\documentclass[10pt,twocolumn]{elsart5p}
\usepackage{graphicx}
\usepackage{dcolumn}
\usepackage{amsmath}
\usepackage{amssymb}
\usepackage{bm}
\usepackage{natbib}
\usepackage{color}

\begin{document}

\begin{frontmatter}

\title{SPIKY: A graphical user interface for monitoring spike train synchrony}

\author{Thomas Kreuz}\ead{thomas.kreuz@cnr.it},
\author{Mario Mulansky} and
\author{Nebojsa Bozanic}

\address{Institute for complex systems, CNR, Sesto Fiorentino, Italy}

\date{\today}

\begin{abstract}
Techniques for recording large-scale neuronal spiking activity are developing very fast. This leads to an increasing demand for algorithms capable of analyzing large amounts of experimental spike train data. One of the most crucial and demanding tasks is the identification of similarity patterns with a very high temporal resolution and across different spatial scales. To address this task, in recent years three time-resolved measures of spike train synchrony have been proposed, the ISI-distance, the SPIKE-distance, and event synchronization. The Matlab source codes for calculating and visualizing these measures have been made publicly available. However, due to the many different possible representations of the results the use of these codes is rather complicated and their application requires some basic knowledge of Matlab. Thus it became desirable to provide a more user-friendly and interactive interface. Here we address this need and present SPIKY, a graphical user interface which facilitates the application of time-resolved measures of spike train synchrony to both simulated and real data. SPIKY includes implementations of the ISI-distance, the SPIKE-distance and SPIKE-synchronization (an improved and simplified extension of event synchronization) which have been optimized with respect to computation speed and memory demand. It also comprises a spike train generator and an event detector which makes it capable of analyzing continuous data. Finally, the SPIKY package includes additional complementary programs aimed at the analysis of large numbers of datasets and the estimation of significance levels.
\end{abstract}


\end{frontmatter}

\newcommand{\abb}{\small\sf}

%
%

\section{\label{s:Intro} Introduction}

Spike train distances are measures of the degree of synchrony between spike trains which yield low values for very similar and high values for very dissimilar spike trains. They are applied in two major scenarios: simultaneous and successive recordings.

The first scenario is the \textit{simultaneous} recording of a neuronal population, typically in a spatial multi-channel setup. If different neurons emit spikes at the same time, these spikes are truly `synchronous' (Greek: `occurring at the same time'). Synchronization between individual neurons has been proven to be of high prevalence in many different neuronal circuits \citep{Tiesinga08, Shlens08}. As of now, many open questions remain regarding the spatial scale and the nature of interactions (pairwise or higher order, see \citeauthor{Nirenberg07}, \citeyear{Nirenberg07}) as well as their functional significance for neuronal coding and information processing \citep{Kumar10}.

In the second scenario the neuronal spiking response is recorded in different time intervals. In order to allow a meaningful comparison there has to be a temporal reference point which is typically set by some kind of trigger (e.g., the onset of an external stimulation). There are two prominent applications for this \textit{successive} trials scenario. Repeated presentation of the same stimulus addresses the reliability of individual neurons \citep{Mainen95}, while different stimuli are used to investigate neuronal coding and to find the features of the response that provide the optimal discrimination (e.g., \citeauthor{Victor05}, \citeyear{Victor05}, for a more general introduction to neural coding cf. \citeauthor{QuianQuiroga13}, \citeyear{QuianQuiroga13}). These two applications are related since for a good clustering performance one needs both a pronounced discrimination between stimuli (high inter-stimulus spike train distances) and a high reliability for the same stimulus (low intra-stimulus spike train distances).

Electrophysiology and other modern recording techniques are developing fast. For both simultaneous population and successive trial recordings they often provide more data than available methods of spike train analysis can handle. There is a lack of algorithms able to identify multiple spike train patterns across different spatial scales and with a high temporal resolution. This is noticeable in both scenarios. In epilepsy, the analysis of the varying similarity patterns of \textit{simultaneously} recorded ensembles of neurons can lead to a better understanding of the mechanisms of seizure generation, propagation, and termination \citep{Truccolo11, Bower12}. Similarly, the analysis of neuronal responses to \textit{successive} presentations of time-dependent stimuli will help to understand the relevance of synchronous firing in neural coding \citep{Miller08}. Moreover, in population recordings it would be even more advantageous to be able to monitor spike train synchrony in realtime. This would be a necessary condition for a prospective epileptic seizure prediction algorithm \citep{Mormann07}, but it could also be very useful for the rapid online decoding needed to control prosthetics \citep{Hochberg06, Sanchez08}.

In recent years, two such time-resolved measures have been proposed: The ISI-distance \citep{Kreuz07c} and the SPIKE-distance \citep{Kreuz13} both rely on instantaneous estimates of spike train dissimilarity which makes it possible to track changes in instantaneous clustering, i.e., time-localized patterns of (dis)similarity among multiple spike trains. Additionally, both measures are parameter-free and time-scale independent. Furthermore, the SPIKE-distance also comes in a causal variant \citep{Kreuz13} which is defined such that the instantaneous values of dissimilarity are derived from past information only so that time-resolved spike train synchrony can be estimated in real-time. Both measures have already been widely used in various contexts (e.g., for the most recent measure, the SPIKE-distance: \citeauthor{Papoutsi13}, \citeyear{Papoutsi13}; \citeauthor{DiPoppa13}, \citeyear{DiPoppa13}; \citeauthor{Sacre14}, \citeyear{Sacre14}). Another time-scale independent and time-resolved method is event synchronization \citep{QuianQuiroga02b}, a sophisticated coincidence detector which quantifies the level of synchrony from the number of quasi-simultaneous appearances of spikes. Originally, it was proposed and used in a bivariate context only. In this paper it is adapted to the time-resolved SPIKY-framework and extended to the multivariate case. Since this involves substantial changes of the original event synchronization, to avoid confusion we term the new, modified measure SPIKE-Synchronization.

With all of these measures spike trains can be analyzed on different spatial and temporal scales, accordingly there are several levels of information extraction \citep{Kreuz12}. In the most detailed representation one instantaneous value is obtained for each pair of spike train. The most condensed representation successive temporal and spatial averaging leads to one single distance value that describes the overall level of synchrony for a group of spike trains over a given time interval. In between these two extremes are spatial averages (dissimilarity profiles) and temporal averages (pairwise dissimilarity matrices). This variety of representations makes a straightforward implementation of the measures in one simple program/function unfeasible. Other important goals are high computational speed, efficient memory management, and applicability to large datasets. What is needed is an intuitive and interactive tool for analyzing spike train data which is able to overcome all of these challenges. 

Here we address this need and present the graphical user interface SPIKY. Given a set of real or simulated spike train data (importable from many different formats), SPIKY calculates the measures of choice and allows the user to switch between many different visualizations such as measure profiles, pairwise dissimilarity matrices, or hierarchical cluster trees. SPIKY also includes the possibility of generating movies which are very useful in order to track the varying patterns of (dis)similarity. SPIKY has been optimized with respect to both computation speed (by using MEX-files, i.e. C-based Matlab executables) and memory demand (by taking advantage of the piecewise linear nature of the dissimilarity profiles). Finally, the SPIKY-package includes two complementary programs. The first program SPIKY\_loop is meant to be used for the analysis of a large number of datasets. The second program SPIKY\_loop\_surro is designed to evaluate the statistical significance of the results obtained for the original dataset by comparing them against the results of spike train surrogates generated from that dataset.

The remainder of this paper is organized as follows. In Section \ref{s:Measures} we present the different measures available in SPIKY and provide some details about their implementation. These measures include the ISI-distance and the SPIKE-distance as well as the latter's realtime variant, and, introduced here, its forward variant (Section \ref{ss:ISI-SPIKE-Distance}). For the SPIKE-distance we propose a correction of the edge-effect (spurious decrease to zero due to auxiliary spikes). In Section \ref{ss:SPIKE-Synchronization} we introduce SPIKE-synchronization, the modified and extended variant of event synchronization. Some improvements realized in the new implementation of the measures are presented in Section \ref{ss:Comparison}. In the same Section we also compare the performance of this new implementation with the one of previously published source codes. In Section \ref{s:Information-extraction} an overview of the different ways to extract information is given. 

SPIKY, our graphical user interface for monitoring spike train synchrony, is presented in Section \ref{s:SPIKY}. In Section \ref{ss:Access} we explain how to get access to the GUI and its documentation. Subsequently, in Section \ref{ss:Structure} we introduce the structure and the workflow of SPIKY, in particular we show how to input spike train data, how to change the layout of the figures and how to export results. The two complementary programs SPIKY\_loop and SPIKY\_loop\_surro are introduced in Sections \ref{ss:GUI-vs-loop} and \ref{ss:Spike-train-surrogates}, respectively. Finally, in Section \ref{s:Discussion} we summarize the methods and the program and present an outlook on future developments.
 
%
%
\section{\label{s:Measures} Measures and implementation}

SPIKY implements four time-resolved measures, one is multivariate and three are bivariate. The multivariate measure, included for comparison, is the standard Peri-Stimulus Time Histogram (PSTH) which measures the overall firing rate. In this Section we give an overview over the three bivariate measures, for more detailed illustrations please refer to the original publications. The ISI-distance \citep{Kreuz07c}, the SPIKE-distance \citep{Kreuz13}, and the here newly proposed SPIKE-synchronization (based on event synchronization, \citeauthor{QuianQuiroga02b}, \citeyear{QuianQuiroga02b}) share several properties, however, there are also a few conceptual differences between the ISI- and the SPIKE-distance on the one hand, and SPIKE-synchronization on the other hand.

All three measures rely on instantaneous values which are normalized between zero and one. The same holds true for the respective temporal averages, the distance values $D_I$ and $D_S$ and the SPIKE-synchronization $S_C$. However, while the two distances are measures of dissimilarity which yield the value zero for identical spike trains, SPIKE-synchronization is a measure of similarity with high values denoting similar spike trains.

While all three measures are time-resolved, the ISI-distance and the SPIKE-distance even have a continuous domain since there exists a unique definition of an instantaneous value ($I (t)$ and $S (t)$, respectively) for every single time instant. The resulting dissimilarity profiles are either piecewise constant (ISI-distance) or piecewise linear (SPIKE-distance). SPIKE-synchronization is time-resolved as well, however, its domain is discrete since instantaneous values $C (t_k)$ are only defined at the times of the spikes. Incidentally, the same distinction holds true regarding the range of values that can be obtained for the measures: it is continuous for the two distances and discrete for SPIKE-synchronization.

All three measures can also be applied to more than two spike trains (spike train number $N > 2$). For the ISI- and the SPIKE-distance this extension is simply the average over all bivariate distances. Extending SPIKE-synchronization is even more straightforward. Essentially, the same spike-based definition holds for both the bivariate and the multivariate case.

\subsection{\label{ss:ISI-SPIKE-Distance} The ISI- and the SPIKE-distance}

The first step in the calculation of the ISI- and the SPIKE-distance is to transform the sequences of discrete spike times $\{t_i^{(1)}\}, i=1,...,M_1$ and $\{t_j^{(2)}\}, j=1,...,M_2$ (with $M_n$ denoting the number of spikes for spike train $n$ with $n = 1,2$) into dissimilarity profiles $I (t)$ and $S (t)$, respectively. Averaging over time yields the respective distance value.

The multivariate extension is obtained as the average over all bivariate distances. Since this average over all pairs of spike trains commutes with the average over time, it is possible to achieve the same kind of time-resolved visualization as in the bivariate case by first calculating the instantaneous average $S^{\mathrm {a}} (t)$ (here for the SPIKE-distance) over all pairwise instantaneous values $S^{mn} (t)$,
\begin{equation} \label{eq:Bivariate-Average}
    S^{\mathrm {a}} (t) = \frac{2}{N(N-1)}\sum_{m=1}^{N-1} \sum_{n=m+1}^N S^{mn} (t).
\end{equation}

The dissimilarity profiles of both measures are based on three piecewise constant quantities (see Fig. \ref{fig:ISI-SPIKE-ES-Illustration}A). These are the time of the preceding spike
\begin{equation} \label{eq:Prev-Spike}
    t_{\mathrm {P}}^{(n)} (t) = \max (t_i^{(n)} | t_i^{(n)} \leq t)  \quad t_1^{(n)} \leq t \leq t_{M_n}^{(n)},
\end{equation}
the time of the following spike
\begin{equation} \label{eq:Foll-Spike}
    t_{\mathrm {F}}^{(n)} (t) = \min (t_i^{(n)} | t_i^{(n)} > t)  \quad t_1^{(n)} \leq t \leq t_{M_n}^{(n)},
\end{equation}
as well as the interspike interval
\begin{equation} \label{eq:ISI}
    x_{\mathrm {ISI}}^{(n)} (t) = t_{\mathrm {F}}^{(n)} (t) - t_{\mathrm {P}}^{(n)} (t).
\end{equation}
The ambiguity regarding the definition of the very first and the very last interspike interval is resolved by placing for each spike train auxiliary leading spikes at time $t = 0$ and auxiliary trailing spikes at time $t = T$ (but see also Section \ref{sss:SPIKE-Distance}).

\subsubsection{\label{sss:ISI-Distance} The ISI-distance}

The ISI-distance, proposed as a bivariate measure in \citet{Kreuz07c} and extended to the multi spike train case in \citet{Kreuz09} was the first spike train distance directly defined as the temporal average of an instantaneous dissimilarity profile. This profile is calculated as the absolute value of the instantaneous ratio between the interspike intervals $x_{\mathrm {ISI}}^{(1)}$ and $x_{\mathrm {ISI}}^{(2)}$ (see Fig. \ref{fig:ISI-SPIKE-ES-Illustration}A) according to:
\begin{equation} \label{eq:ISI-Ratio}
    I (t) = \frac{|x_{\mathrm {ISI}}^{(1)} (t) - x_{\mathrm {ISI}}^{(2)} (t)|}{\max (x_{\mathrm {ISI}}^{(1)} (t), x_{\mathrm {ISI}}^{(2)} (t))}
\end{equation}
Since the ISI-values only change at the times of spikes, the dissimilarity profile is piecewise constant (with discontinuities at the spikes). The ISI-ratio equals zero for identical interspike intervals in the two spike trains, and approaches one in intervals in which one spike train is much faster than the other. The ISI-distance is defined as the temporal average of this absolute ISI-ratio:
\begin{equation} \label{eq:Temporal-Average}
    D_I = \frac{1}{T} \int_0^T dt I (t).
\end{equation}

\subsubsection{\label{sss:SPIKE-Distance} The SPIKE-distance}

The SPIKE-distance (see \citeauthor{Kreuz11}, \citeyear{Kreuz11}, for the original proposal and \citeauthor{Kreuz13}, \citeyear{Kreuz13}; \citeauthor{Kreuz12}, \citeyear{Kreuz12}, for the definite version presented here) is the centerpiece of SPIKY. In contrast to the ISI-distance, it considers the exact timing of the spikes. 
%
%
\begin{figure}
    \includegraphics[width=85mm]{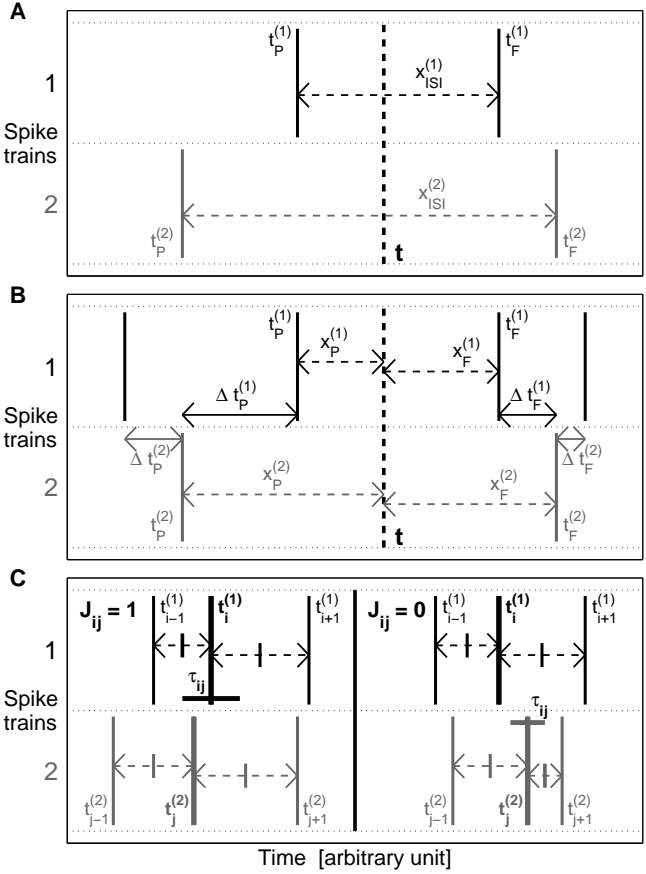}
    \caption{\abb\label{fig:ISI-SPIKE-ES-Illustration} A. ISI-distance. Illustration of the local quantities used to define the dissimilarity profile $I (t)$ for an arbitrary time instant $t$.   B. SPIKE-distance. Illustration of the additional local quantities needed for the calculation of the dissimilarity profile $S (t)$.   C. SPIKE-synchronization. Illustration of the adaptive coincidence detection (which was originally proposed for event synchronization). While in the first half the middle spikes $t_i^{(1)}$ and $t_j^{(2)}$} are coincident, the middle spikes in the second half are not.
\end{figure}
%

The dissimilarity profile is calculated in two steps: First for each spike a spike time difference is calculated, then for each time instant the relevant spike time differences are selected, weighted, and normalized. Here `relevant' means local; each time instant is surrounded by four corner spikes: the preceding spike from the first spike train $t_{\mathrm {P}}^{(1)}$, the following spike from the first spike train $t_{\mathrm {F}}^{(1)}$, the preceding spike from the second spike train $t_{\mathrm {P}}^{(2)}$, and, finally, the following spike from the second spike train $t_{\mathrm {F}}^{(2)}$. To each of these corner spikes one assigns the distance to the nearest spike in the other spike train, for example, for the previous spike of the first spike train:
\begin{equation} \label{eq:Delta-Corner-Spike}
     \Delta t_{\mathrm {P}}^{(1)} (t) = \min_i (| t_{\mathrm {P}}^{(1)} (t) - t_i^{(2)} |)
\end{equation}
and analogously for $t_{\mathrm {F}}^{(1)}$, $t_{\mathrm {P}}^{(2)}$, and $t_{\mathrm {F}}^{(2)}$  (see Fig. \ref{fig:ISI-SPIKE-ES-Illustration}B). Subsequently, for each spike train separately, a locally weighted average is employed such that the differences for the closer spike dominate; for each spike train $n = 1, 2$ the weighting factors are the intervals to the previous and to the following spikes:
\begin{equation} \label{eq:Prev-Spike-Dist}
     x_{\mathrm {P}}^{(n)} (t) = t - t_{\mathrm {P}}^{(n)} (t)
\end{equation}
and
\begin{equation} \label{eq:Foll-Spike-Dist}
     x_{\mathrm {F}}^{(n)} (t) = t_{\mathrm {F}}^{(n)} (t) - t.
\end{equation}
The local weighting for the spike time differences of the first spike train reads:
\begin{equation} \label{eq:Bi-Spike-Diss-First}
     S_1 (t) = \frac{\Delta t_{\mathrm {P}}^{(1)} (t) x_{\mathrm {F}}^{(1)} (t) + \Delta t_{\mathrm {F}}^{(1)} (t) x_{\mathrm {P}}^{(1)} (t)}{x_{\mathrm {ISI}}^{(1)} (t)},
\end{equation}
and analogously $S_2 (t)$ is obtained for the second spike train. Averaging over the two spike train contributions and normalizing by the mean interspike interval yields:
\begin{equation} \label{eq:Bi-Spike-Diss-Intermediate}
     S'' (t) = \frac{S_1 (t) + S_2 (t)}{2 \langle x_{\mathrm {ISI}}^{(n)} (t) \rangle_n}.
\end{equation}
This quantity sums the spike time differences for each spike train weighted according to the relative distance of the corner spike from the time instant under investigation. This way relative distances within each spike train are taken care of, while relative distances between spike trains are not. In order to get these ratios straight and to account for differences in firing rate, in a last step the two contributions from the two spike trains are locally weighted by their instantaneous interspike intervals. This leads to the definition of the dissimilarity profile:
\begin{equation} \label{eq:Bi-Spike-Diss-Improved}
     S (t) = \frac{S_1 (t) x_{\mathrm {ISI}}^{(2)} (t) + S_2 (t) x_{\mathrm {ISI}}^{(1)} (t)}{2 \langle x_{\mathrm {ISI}}^{(n)} (t) \rangle_n^2}.
\end{equation}
Again, the overall distance value is defined as the temporal average of the dissimilarity profile:
\begin{equation} \label{eq:Temporal-Average2}
    D_S = \frac{1}{T} \int_0^T dt S (t).
\end{equation}

Since the dissimilarity profile $S (t)$ is obtained from a linear interpolation of piecewise constant quantities, it is piecewise linear (with potential discontinuities at the spikes). Both the dissimilarity profile $S (t)$ and the SPIKE-distance $D_S$ are bounded in the interval $[0, 1]$. The distance value $D_S = 0$ is obtained for identical spike trains only.

Due to the finite recording time there is an ambiguity regarding the definitions of the initial distance to the preceding spike, the final distance to the following spike, as well as the very first and the very last interspike intervals. In previous implementations of the SPIKE-distance this ambiguity was resolved by adding to each spike train an auxiliary leading spike at time $t = 0$ and an auxiliary trailing spike at time $t = T$. This lead to spurious synchrony at the edges where by construction the dissimilarity profile reached the zero value. Here we partly correct this edge effect by incorporating all the information that is available. We describe the correction only for the beginning of the recording, an analogous procedure is applied at the end of the recording.

We count the auxiliary spikes as normal spikes which can be nearest neighbors to other spikes. But instead of calculating their spike time distance (which is always zero) we use the spike time difference of the first real spike. For the first interspike interval we know that it is at least the distance to the first spike $t_1-t_0 = t_1$ but it could be longer. So to take the local firing rate into consideration we set
\begin{equation} \label{eq:Corrected-First-ISI}
    x_{\mathrm {ISI}} (0) = \max ( t_1, t_2 - t_1 ).
\end{equation}
where we use the length of the first known interspike interval $t_2-t_1$ as a lower limit. In case $t_1$ is smaller than $t_2-t_1$ we get at least a crude estimate of how much longer the first interspike interval could have been.

\subsubsection{\label{sss:Realtime-Spike-Distance} Realtime SPIKE-distance}

In contrast to the dissimilarity profile $S (t)$ of the regular SPIKE-distance, the dissimilarity profile $S_r (t)$ of the realtime SPIKE-distance can be calculated online because it relies on past information only. From the perspective of an online measure, the information provided by the following spikes, both their position and the length of the interspike interval, is not yet available. Like the profile of the regular SPIKE-distance, this causal variant is also based on local spike time differences but now only two corner spikes are available, and the spikes of comparison are restricted to past spikes, e.g., for the preceding spike of the first spike train:
\begin{equation} \label{eq:Delta-Corner-Spike-Realtime}
     \Delta t_{\mathrm {P}}^{(1)} (t) = \min_{i:t_i \leq t} (| t_{\mathrm {P}}^{(1)} (t) - t_i^{(2)} |).
\end{equation}
Since there are no following spikes available, there is no local weighting. There is no interspike interval either, so the normalization is achieved by dividing the average corner spike difference by twice the average time interval to the preceding spikes (Eq. \ref{eq:Prev-Spike-Dist}). This yields a causal indicator of local spike train dissimilarity:
\begin{equation} \label{eq:Bi-Spike-Diss-RT}
    S_r (t) = \frac{ \Delta t_{\mathrm {P}}^{(1)} (t) + \Delta t_{\mathrm {P}}^{(2)} (t)} {4 \langle x_{\mathrm {P}}^{(n)} (t) \rangle_n}.
\end{equation}

\subsubsection{\label{sss:Forward-Spike-Distance} Forward SPIKE-distance}

The dissimilarity profile $S_f (t)$ of the forward SPIKE-distance is `inverse' to the profile of the realtime SPIKE-distance. Instead of relying on past information only it relies on forward information only. It can be used in triggered temporal averaging in order to evaluate the (causal) effect of certain spikes or of specific stimuli features on future spiking. Again, for each time instant there are just two corner spikes and the potential nearest spikes in the other spike train are future spikes only. Thus, the spike time difference for the following spike of the first spike train reads:
\begin{equation} \label{eq:Delta-Corner-Spike-Forward}
     \Delta t_{\mathrm {F}}^{(1)} (t) = \min_{i:t_i \geq t} (| t_{\mathrm {F}}^{(1)} (t) - t_i^{(2)} |)
\end{equation}
and accordingly for the following spike of the second spike train. In analogy to Eq. \ref{eq:Bi-Spike-Diss-RT}, an indicator of local spike train dissimilarity is obtained as follows:
\begin{equation} \label{eq:Bi-Spike-Diss-FT}
    S_f (t) = \frac{ \Delta t_{\mathrm {F}}^{(1)} (t) + \Delta t_{\mathrm {F}}^{(2)} (t)} {4 \langle x_{\mathrm {F}}^{(n)} (t) \rangle_n}.
\end{equation}

\subsection{\label{ss:SPIKE-Synchronization} SPIKE-synchronization}

SPIKE-synchronization quantifies the degree of synchrony from the relative number of quasi-simultaneous appearances of spikes. Since it builds on the same bivariate and adaptive coincidence detection that was used for event synchronization (\citeauthor{QuianQuiroga02b}, \citeyear{QuianQuiroga02b}; see also \citeauthor{Kreuz07c}, \citeyear{Kreuz07c}), SPIKE-synchronization is parameter- and scale-free as well.

Coincidence detection typically uses a coincidence window $\tau_{ij}^{(1,2)}$ which denotes the time lag below which two spikes from two different spike trains, $t_i^{(1)}$ and $t_j^{(2)}$, are considered to be coincident. For both event synchronization and SPIKE synchronization this coincidence window is adapted to the local spike rates (see Fig. \ref{fig:ISI-SPIKE-ES-Illustration}C):
\begin{equation} \label{eq:Coincidence-MaxDist}
    \tau_{ij}^{(1,2)} = \min \{t_{i+1}^{(1)} - t_i^{(1)}, t_i^{(1)} - t_{i-1}^{(1)},
                           t_{j+1}^{(2)} - t_j^{(2)}, t_j^{(2)} - t_{j-1}^{(2)}\}/2.
\end{equation}
The coincidence criterion can be quantified by means of a coincidence indicator
\begin{equation} \label{eq:SPIKE-Coincidence}
    C_i^{(1)} = \begin{cases}
                      1     & {\rm if} ~~ \min_j(|t_i^{(1)} - t_j^{(2)}|) < \tau_{ij}^{(1,2)} \cr
                      0     & {\rm otherwise}
                  \end{cases}
\end{equation}
(and analogously for $C_j^{(2)}$) which assigns to each spike either a one or a zero depending on whether it is part of a coincidence or not. The minimum function in Eq. \ref{eq:SPIKE-Coincidence} takes already into account that a spike can at most be coincident with one spike (the nearest one) in the other spike train. This is a consequence of the adaptive definition of $\tau_{ij}^{(1,2)}$ in Eq. \ref{eq:Coincidence-MaxDist} and the `$<$' in Eq. \ref{eq:SPIKE-Coincidence} (which has been changed from the `$\leq$' used in the original definition of event synchronization). In case a spike is right in the middle between two spikes from the other spike train there is no ambiguity any more since there is no coincidence.

This way we have defined a coincidence indicator for each individual spike of the two spike trains. In order to obtain one combined similarity profile we pool the spikes of the two spike trains as well as their coincidence indicators by introducing one overall spike index $k$. In case there exist exact matches (pairs of perfectly coincident spikes) $k$ counts over both spikes. This yields one unified set of coincidence indicators $C_k$ in which according to Eqs. \ref{eq:Coincidence-MaxDist} and \ref{eq:SPIKE-Coincidence} each coincidence leads to a pair of consecutive ones.

From this discrete set of coincidence indicators $C_k$ the SPIKE-Synchronization profile $C (t_k)$ is obtained via $C (t_k) = C (k)$. Finally, SPIKE-Synchronization is defined as the average value of this profile
\begin{equation} \label{eq:Bi-SPIKE-Synchronization}
   S_C = \frac{1}{M} \sum_{k=1}^M C (t_k)
\end{equation}
with $M = M_1 + M_2$ denoting the total number of spikes in the pooled spike train. The interpretation is very intuitive: $S_C$ quantifies the fraction of all spikes in the two spike trains that are coincident. It is zero for spike trains without any coincidences, and reaches one if and only if the two spike trains consist only of pairs of coincident spikes.

The extension to the case of more than two spike trains ($N > 2$) is straightforward. First, bivariate coincidence detection is performed for each pair of spike trains $(n,m)$. Generalizing Eq. \ref{eq:SPIKE-Coincidence} gives the coincidence indicators
\begin{equation} \label{eq:Bi-SPIKE-Coincidence}
    C_i^{(n,m)} = \begin{cases}
                      1     & {\rm if} ~~ \min_j(|t_i^{(n)} - t_j^{(m)}|) < \tau_{ij}^{(n,m)} \cr
                      0     & {\rm otherwise}
                  \end{cases}
\end{equation}
where $\tau_{ij}^{(n,m)}$ is defined as in Eq. \ref{eq:Coincidence-MaxDist}, but for arbitrary spike trains $n$ and $m$. Subsequently, for each spike of every spike train a normalized coincidence counter
\begin{equation} \label{eq:Multi-SPIKE-Coincidence}
 C_i^{(n)} = \frac1{N-1}\sum_{m\neq n} C_i^{(n,m)}.
\end{equation}
is obtained by averaging over all $N-1$ bivariate coincidence indicators involving the spike train $n$.

As in the bivariate case, pooling leads to just one set of normalized coincidence counters $C_k$ each of which can obtain any one out of this finite set of values: $0, 1/(N-1), ..., (N-1)/(N-1) = 1$. Again, the multivariate SPIKE-Synchronization profile $C (t_k)$ is obtained via $C (t_k) = C (k)$ and its average value yields the multivariate SPIKE-Synchronization
\begin{equation} \label{eq:Multi-SPIKE-Synchronization}
   S_C = \frac{1}{M} \sum_{k=1}^M C (t_k)
\end{equation}
where $M = \sum_n^N M_n$ again denotes the overall number of spikes.

Note that Eq. \ref{eq:Multi-SPIKE-Synchronization} is completely analogous to Eq. \ref{eq:Bi-SPIKE-Synchronization}. Moreover, setting $n=1$, $m=2$, and $N=2$ in Eqs. \ref{eq:Bi-SPIKE-Coincidence} and \ref{eq:Multi-SPIKE-Coincidence} retrieves Eq. \ref{eq:SPIKE-Coincidence}. Therefore, the bivariate equations are just a special case of the more general multivariate formulation.

\begin{figure}
    \includegraphics[width=85mm]{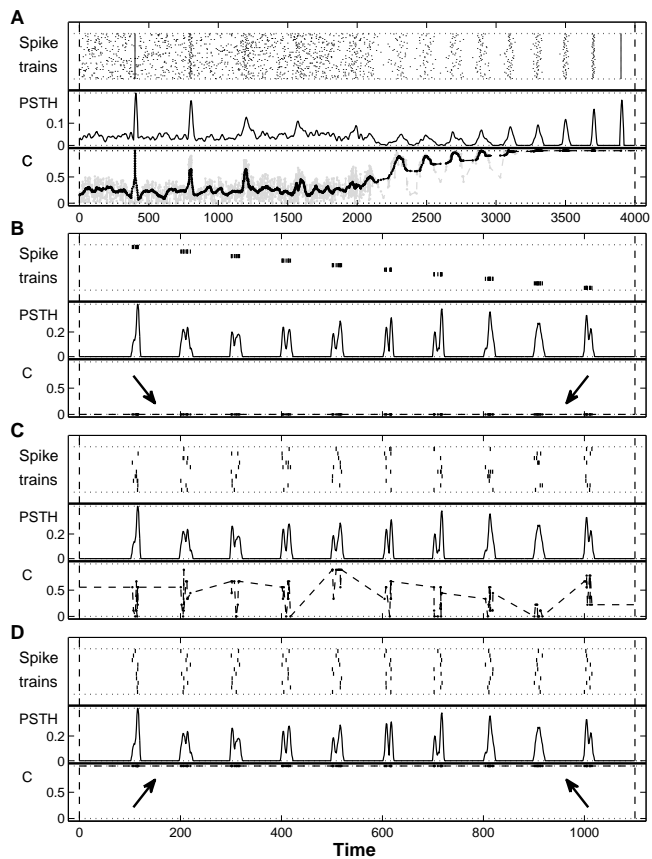}
    \caption{\abb\label{fig:PSTH-ES} Comparison of PSTH and SPIKE-synchronization. For the latter we added a dashed similarity profile which serves as a visual aid only.   A. Multivariate example with $50$ spike trains. In the first half within the noisy background there are $4$ regularly spaced spiking events with increasing jitter. The second half consists of $10$ spiking events with decreasing jitter but now without any noisy background. In the noisy first half PSTH and the smoothed SPIKE-synchronization exhibit very similar profiles. The fact that the firing events become more distinct in the second half is indicated by the smoothed SPIKE-synchronization as a gradual increase to synchronization. In the PSTH the peaks become more and more narrow.   B-D. By construction the pooled spike train of these examples is identical consisting of $10$ evenly spaced bursts. The only difference is the distribution of the spikes among the individual spike trains which varies from low via intermediate to high synchrony. Whereas the PSTH is the same for all three examples, SPIKE-synchronization correctly indicates the increase in synchrony (note that in subplot B SPIKE-synchronization attains the value zero and in subplot D the value one over the whole time interval, see arrows).   B. Synfire chain of bursts.   C. Random distribution of spikes among spike trains.   D. High reliability. Each spike train contains one spike per firing event.}
\end{figure}

Accordingly, the interpretation of SPIKE synchronization as the overall fraction of potential coincidences that are actually realized is general and holds for both bivariate and multivariate datasets. SPIKE-Synchronization is zero if and only if the spike trains do not contain any coincidences, and reaches one if and only if each spike in every spike train has one matching spike in all the other spike trains. Examples for both of these extreme cases can be found in subplots B and D of Fig. \ref{fig:PSTH-ES}.

%
%
%
In contrast to $I (t)$ and $S (t)$, the time-resolved SPIKE-synchronization $C (t_k)$ is a measure of similarity. The Peri-Stimulus Time Histogram is oriented the same way so it makes sense to compare these two measures (Fig. \ref{fig:PSTH-ES}). The SPIKE-Synchronization profile $C (t_k)$ is only defined at the times of the spikes but a better visualization can be achieved by connecting the individual dots. For larger spike train datasets (here example A) it also makes sense to smooth the profile with a moving average of appropriate order. Fig. \ref{fig:PSTH-ES}A demonstrates the similarities and dissimilarities between the PSTH and SPIKE-synchronization on a rather general example. Fig. \ref{fig:PSTH-ES}B-D shows that SPIKE-synchronization, in contrast to  the PSTH, is a true measure of spike train synchrony (see also \citeauthor{Kreuz11b}, \citeyear{Kreuz11b}). Since the PSTH is invariant to shuffling spikes among the spike trains, it yields the same value regardless of how spikes are distributed among the different spike trains.

\subsection{\label{ss:Comparison} Comparison with other implementations}

The very first all-in-one implementation of the ISI- and the SPIKE-distance\footnote{This Section concerns only the ISI- and the SPIKE-distance. Since SPIKE-Synchronization is a new proposal there are no other implementations.} was published online at the time of the publication of \cite{Kreuz13}. In between this first and our current release, several other source codes written in various languages and for different platforms have been made available. The most prominent examples are the Python-Implementation of the SPIKE-distance courtesy of Jeremy Fix and available on his homepage\footnote{http://jeremy.fix.free.fr/Softwares/spike.html}, and the C++-Implementation of both ISI- and SPIKE-distance courtesy of R{\v a}zvan Florian and hosted on GitHub\footnote{https://github.com/modulus-metric/spike-train-metrics}. The SPIKE-distance was also implemented in the commercially distributed HRLAnalysis$^{TM}$ software suite \citep{Thibeault14} designed for the analysis of large-scale spiking neural data. Note that all of these implementations are restricted to the dissimilarity profile and its temporal average (the overall dissimilarity). In contrast, SPIKY also allows the user to interactively access all the other different ways to extract information that will be introduced in Section \ref{s:Information-extraction}.

All of these codes for calculating the ISI- and the SPIKE-distance rely on equidistantly sampled dissimilarity profiles, and the same holds true for codes of event synchronization. Typically the precision is set to the sampling interval of the neuronal recording. Since the dissimilarity profiles have to be calculated and stored for each pair of spike trains, one obtains, for each measure, a matrix of order `number of sampled time instants' $\times$ `number of spike train pairs' (i.e., $\# (t_s) \times N(N-1)/2$). For small sampling intervals and large numbers of spike trains this leads to memory problems.

\begin{figure}
    \includegraphics[width=85mm]{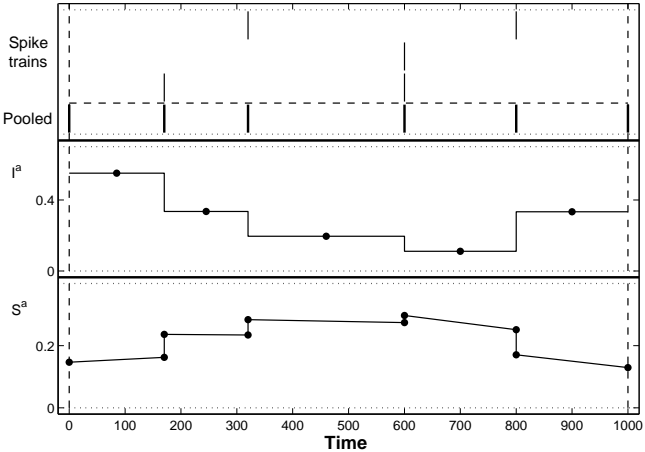}
    \caption{\abb\label{fig:No-sampling} Illustration of the memory efficient storage of the dissimilarity profile for both the ISI- and the SPIKE-distance.}
\end{figure}

In SPIKY we use an optimized and more memory-efficient way of storing the results. We make use of the fact that the dissimilarity profile $I (t)$ of the ISI-distance is piecewise constant and the dissimilarity profile $S (t)$ of the SPIKE-distance is piecewise linear. Each constant/linear interval runs from one spike of the pooled spike train to the next. Thus, for each such interval (and for each pair of spike trains) we have to store only one value for the ISI-dissimilarity and two values for the SPIKE-dissimilarity, one at the beginning and one at the end of the interval (see Fig. \ref{fig:No-sampling}). The memory gain is proportional to the number of sample points per interspike interval in the pooled spike train and is typically much larger than one.
%
%
%

The dissimilarity profiles exhibit instantaneous jumps at the times of the spikes since this is where the lengths of the interspike intervals and the identity of the previous and the following spikes change abruptly or where a new coincidence is counted. For sampled dissimilarity profiles one has to `cut the corners' of these instantaneous jumps. This leads to an estimation error which increases with the sampling interval. In contrast, within the new implementation each spike marks both the end of the previous and the beginning of the next interval and it becomes possible to store two dissimilarity values for these points. This way the integration from one spike of the pooled spike train to the next can be performed over the full interval. Thus, besides being far more memory-efficient, the new implementation also computes the exact distance values without any spurious dependence on the sampling interval.

The third effect of the new implementation is a considerable speed-up. To show this (and to provide relative computational costs for the different measures) we here calculate the speed gain achieved by going from the old equidistantly sampled implementation to the new minimally sampled implementation of the ISI- and the SPIKE-distance. 

\begin{figure}
    \includegraphics[width=85mm]{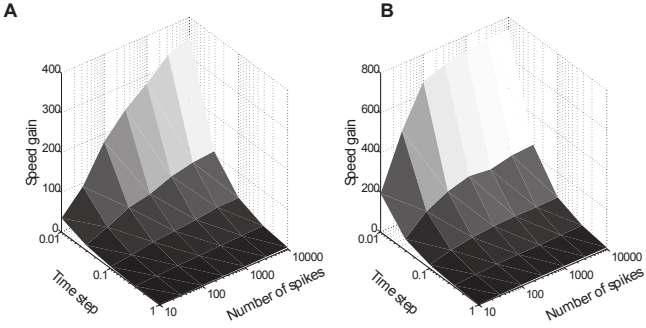}
        \caption{\abb\label{fig:Performance-Comparison} Speed gain achieved by the new (minimally sampled) implementation of A. ISI-distance and B. SPIKE-distance with respect to the old (equidistantly sampled) implementation in dependence on the number of spikes and on the time step used by the old implementation.}
\end{figure}

As benchmark we use the comprehensive performance comparison carried out by \cite{Rusu14}. In this test the authors compared the performance of their newly proposed modulus-metric with the performance of previously proposed spike train distances including the ISI- and the SPIKE-distance. Like them we used two random spike trains with different numbers of spikes. However, since we were also interested in applications to larger datasets, we extended the maximum number of spikes from $500$ to $10000$ spikes. The firing rate is kept constant, hence the duration of the trial increases with the number of spikes. For a fair comparison we implemented all algorithms in C++ and ran them on an Intel i7-4700MQ CPU @ $2.4$ GHz. All speed gains were averaged over $10,000$ trials (Fig. \ref{fig:Performance-Comparison}).

In a first step we replicated the results of \citet{Rusu14} who had calculated the ISI- and SPIKE-distances using dissimilarity profiles that were equidistantly sampled with a fixed time step of dt = $1$ ms. Rerunning the same algorithms on our computer, we could reproduce their results (for less than $500$ spikes) within the same order of magnitude. Subsequently, we measured the running times using minimally sampled dissimilarity profiles and found that for both the ISI- and the SPIKE-distance the new implementation was considerably faster than the old implementation. As expected, the speed gain depends critically on the sampling rate used for the old implementation and is particularly large for densely sampled data (Fig. \ref{fig:Performance-Comparison}).

Apart from minimal sampling, SPIKY includes another improvement which strongly increases the overall performance. The source codes published along with \cite{Kreuz13} were entirely written in Matlab. In contrast, the new SPIKY-implementation uses C-based Matlab executables (MEX)-files for the most time-consuming parts and this leads to another enormous performance boost (typically by a factor between $3$ and $30$).
%
%
%

%
\section{\label{s:Information-extraction} Levels of information extraction}

The ISI- and the SPIKE-distance combine a variety of properties that make them well suited for the application to real data. In particular, they are conceptually simple, computationally efficient, and easy to visualize in a time-resolved manner. By taking into account only the preceding and the following spike in each spike train, these distances rely on local information only. They are also time-scale-adaptive since the information used is not contained within a window of fixed size but rather within a time frame whose size depends on the local rate of each spike train. 

\begin{figure}
    \includegraphics[width=85mm]{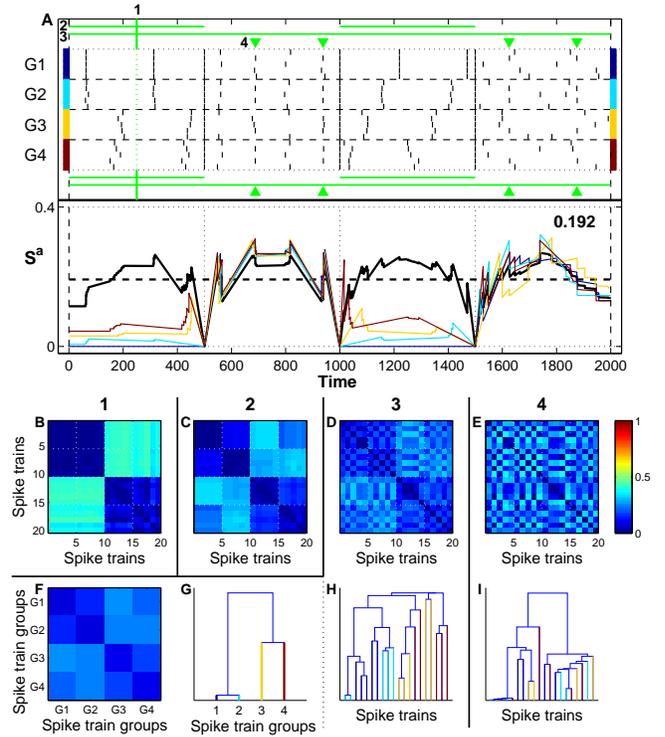}
    \caption{\abb\label{fig:SPIKE-Representations} The different levels of information extraction for the SPIKE-distance.   A. Top: Spike raster plot of $20$ artificially generated spike trains divided in $4$ spike train groups of $5$ spike trains each. The clustering behavior changes every $500$ ms. Bottom: Dissimilarity profiles of the SPIKE-distance for the four spike train groups (thin color-coded lines) and for all spike trains (thick black line). The overall dissimilarity is defined as the temporal average of the dissimilarity profile of all spike trains ($0.192$ in this case) and is marked by a dashed horizontal line. The green lines and symbols on top of the raster plot mark temporal instants and intervals results of which are detailed in the subplots below.  B-E. Matrices of pairwise instantaneous dissimilarity values for a single time instant (mark 1, subplot B), for two selective averages (over non-consecutive intervals in mark 2, subplot C, and over the whole interval in mark 3, subplot D) and for a triggered average (mark 4, subplot E).  F-G. For the overall average (mark 3) we also show the matrices of overall pairwise instantaneous dissimilarity values for the $4$ spike train groups (F) and the corresponding dendrogram (G).  H-I. Dendrogram of spike train matrices in D and E. Note that in contrast to the overall average (mark 3, subplot H) the triggered average (mark 4, subplot I) captures the local similarity between $5$ of the spike trains.}
\end{figure}

Moreover, the sensitivity to spike timing and the instantaneous reliability achieved by the SPIKE-distance opens up many new possibilities in multi-neuron spike train analysis \citep{Kreuz13}. These build upon the fact that there are several ways to extract information all of which we describe in the following. As an illustration we use the detailed analysis of an artificially generated spike train dataset with the SPIKE-distance (for the raster plot see upper part of Fig. \ref{fig:SPIKE-Representations}A).

Since the profile of SPIKE-synchronization is only defined at the times of the spikes and not at any instantaneous instants, for this measure only a few of these levels apply, as we will detail below.

%
%
%

\subsection{\label{ss:Full-matrix-and-cross-sections} Full matrix and cross sections}

The starting point is the most detailed representation in which one instantaneous value is obtained for each pair of spike trains (see Eq. \ref{eq:Bi-Spike-Diss-Improved}). This representation could be viewed as a movie of a symmetric pairwise dissimilarity matrix in which each frame corresponds to one time instant (an example can be found in the supplementary material of \citeauthor{Kreuz13}, \citeyear{Kreuz13}). For a movie of finite length the time axis necessarily has to be sampled but in principle this most detailed representation consists of an infinite number of values. However, since all dissimilarity profiles are piecewise linear there is a lot of redundancy. 

One step towards a more compact and memory-efficient representation is to store all pairwise dissimilarity profiles in a matrix of size `number of interspike intervals in the pooled spike train' $\times$ `number of spike train pairs' ($\times 2$ for the SPIKE-distance, see Section \ref{ss:Comparison}). From this two-dimensional matrix it is possible to extract both kinds of cross sections. By selecting a pair of spike trains, one obtains the bivariate dissimilarity profile $S (t)$ for this pair of spike trains. Selecting a time instant $t_s$ (and using linear interpolation for time instants in between spikes) yields an instantaneous matrix of pairwise spike train dissimilarities $S_{mn}(t_s)$ (see Fig. \ref{fig:SPIKE-Representations}B). This matrix can be used to divide the spike trains into instantaneous clusters, that is, groups of spike trains with low intra-group and high inter-group dissimilarity.

For SPIKE-synchronization it is possible to select pairwise dissimilarity profiles but instantaneous matrices are not defined.

\subsection{\label{ss:Spatial-and-temporal-Averaging} Spatial and temporal averaging}

Another way to reduce the information of the dissimilarity matrix is averaging. There are two possibilities that commute: the spatial average over spike train pairs and the temporal average. Since the spatial average over spike train pairs can be done locally it yields a dissimilarity profile for the whole population. Examples for averages over four different spike train groups as well as over all spike trains are shown in the lower subplot of Fig. \ref{fig:SPIKE-Representations}A. Temporal averaging over certain intervals on the other hand leads to a bivariate distance matrix (see Fig. \ref{fig:SPIKE-Representations}C and D for examples of non-continuous and continuous intervals). In real data, these temporal intervals could be chosen to correspond to different external conditions such as normal vs. pathological, asleep vs. awake, target vs. non-target stimulus, or presence/absence of a certain channel blocker. 

A combination of temporal and spatial averaging can be seen in Fig. \ref{fig:SPIKE-Representations}F. This dissimilarity matrix is obtained from the overall temporal average shown in Fig. \ref{fig:SPIKE-Representations}D by (spatially) averaging over the $16$ submatrices and thus depicts the pairwise spike train group dissimilarity ($4 \times 4$ instead of $20 \times 20$). Fig. \ref{fig:SPIKE-Representations}G shows the respective dendrogram. In applications to real data, these groups could be different neuronal populations or responses to different stimuli, depending on whether the spike trains were recorded simultaneously or successively. Finally, successive application of spatial average over all spike train pairs and temporal average over the whole interval results in just one distance value that describes the overall level of dissimilarity for the entire dataset. In Fig. \ref{fig:SPIKE-Representations}A this value is stated in the upper right of the lower subplot.

Both spatial and temporal averaging are well-defined for SPIKE-synchronization, and so is the one overall value.

\subsection{\label{ss:Triggered-Averaging} Triggered averaging}

The fact that there are no limits to the temporal resolution allows further analyses such as internally or externally triggered temporal averaging. Here, the matrices are averaged over certain trigger time instants only. The idea is to check whether this triggered temporal average is significantly different from the global average since this would indicate that something peculiar is happening at these trigger instants.

The trigger times can either be obtained from external influences (such as the occurrence of certain features in a stimulus) or from internal conditions (such as the spike times of a certain spike train). External triggering is a standard tool to address questions of neural coding, for example, it can be used to evaluate the influence of localized stimulus features on the reliability of neurons under repeated stimulation. In multi-neuron data, internal triggering might help to uncover the connectivity in neural networks or to detect converging or diverging patterns of firing propagation.

An example is shown in Fig. \ref{fig:SPIKE-Representations}E. Here, neurons $2$, $9$, $14$, and $19$ follow the $7$-th neuron during the 2nd and the 4th $500$-ms subinterval. This can be revealed by triggering on the spike times of neuron $7$ during these subintervals. The difference between the dissimilarity matrix of the full interval in Fig. \ref{fig:SPIKE-Representations}D and the triggered dissimilarity matrix of Fig. \ref{fig:SPIKE-Representations}E can be best seen by comparing the respective dendrograms in Fig. \ref{fig:SPIKE-Representations}H and \ref{fig:SPIKE-Representations}I.

Since instantaneous values are not defined, triggered averaging does not make any sense for SPIKE-synchronization.

%
%
\section{\label{s:SPIKY} SPIKY}

SPIKY is a graphical user interface for monitoring synchrony between artificially simulated or experimentally recorded neuronal spike trains. It contains implementations of the ISI-distance, the SPIKE-distance, and SPIKE-synchronization. Moreover, SPIKY includes the forward variant of the SPIKE-distance as well as a `simulation' of the realtime SPIKE-distance. The latter calculates the values the realtime version would give but instead of going forward in real time it just calculates all values in parallel (see also Section \ref{ss:Outlook}). The source codes are written in Matlab (MathWorks Inc, Natick, MA, USA) with the most time-consuming loops coded in MEX-files\footnote{In our case these are subroutines written in C. However, as some users may not have access to a suitable C compiler, SPIKY contains the (slower) pure Matlab code as well.}. Consequently, SPIKY is not stand-alone but requires Matlab to run.

The following Sections give a broad overview of the features of SPIKY. For a more detailed description of the specific procedures that realize these features please refer to the documentation and the webpages listed below.

\subsection{\label{ss:Access} Access to SPIKY and how to get started}

SPIKY is distributed under the BSD licence (Copyright (c) 2014, Thomas Kreuz, Nebojsa Bozanic. All rights reserved.). A zip-package containing all the necessary files can be accessed for free on the download page\footnote{http://www.fi.isc.cnr.it/users/thomas.kreuz/Source-Code/SPIKY.html}. This package also contains a folder with documentation (such as a FAQ-file and an introduction to all individual elements of SPIKY and all individual files of the SPIKY-package). Further information and many demonstrations (both images and movies) can be found on the download page and on the SPIKY Facebook-page\footnote{https://www.facebook.com/SPIKYgui}. Both of these pages are used to announce updates and distribute the latest information about new features. They also provide an opportunity to give feedback and ask questions. Moreover, the Facebook-page includes various screen recordings with voice-over in which the user is guided step by step through some of the most important features of SPIKY. All of these movies can also be viewed on the SPIKY Youtube-channel\footnote{https://www.youtube.com/user/SPIKYgui1}.

After downloading SPIKY the user has to first extract the zip-package which leaves all files in one folder named `SPIKY'. If the system has a suitable MEX-compiler installed, the MEX-files can be compiled from within this folder by running the m-file `SPIKY\_compile\_MEX'. The program is started with the m-file `SPIKY'.

When SPIKY is running, the user has the option to select to view short hints (`tooltips') when hovering above individual elements of the graphical user interface. An overview of all the information contained in the hints can be found in the documentation file `SPIKY-Elements'. Furthermore, at each step the suggested element for the next user action is highlighted by a bold font.

To get the user started quickly, SPIKY provides a few example datasets from previous publications. The most useful example is the `Clustering' dataset which has already been used in several figures as well as in the supplementary movie of \cite{Kreuz13}. The best way to get acquainted with SPIKY is to advance from panel to panel by pressing the highlighted button. When the end is reached the user can reset and run the same example again while changing some parameters in order to see the consequences. Note that it is not necessary to set all the parameters each time when SPIKY is started. Rather, it is possible to use the file `SPIKY\_f\_user\_interface' to set and modify the spike train data as well as the parameters (again, the dataset `Clustering' provides an example).

\subsection{\label{ss:Structure} Structure and workflow of SPIKY}

\begin{figure}
    \includegraphics[width=85mm]{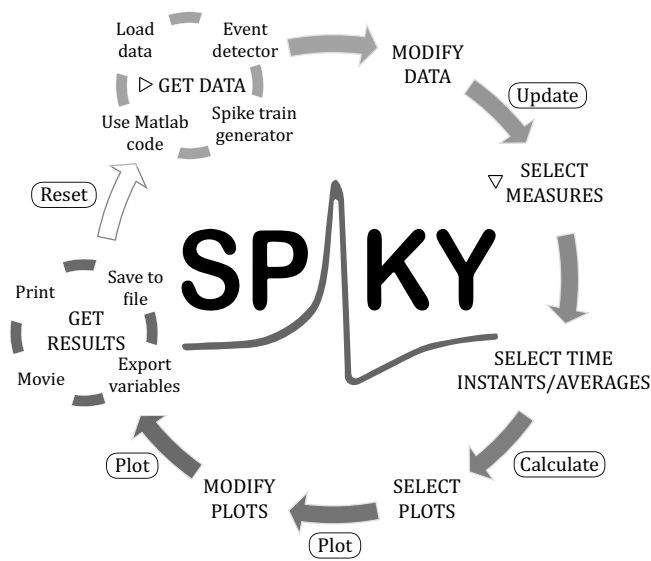}
    \caption{\abb\label{fig:SPIKY-Flowchart} SPIKY-logo (center) and flowchart describing the workflow of SPIKY from the input of spike train data to the output of results. A typical SPIKY-session begins at the top with `Get data', then goes clockwise and ends on the left with `Get results'. For clarity we omitted two possible actions which can be performed from multiple positions on the workflow circle: From any point it is possible to reset to the very beginning (`reset' leads to the symbol $\triangleright$), and from any later point it is possible to jump back before the measure calculation (`reset with the same data' leads to the symbol $\triangledown$).}
\end{figure}

Overall, SPIKY has a rather linear workflow, however, it is much more interactive than previous implementations of the measures and there are many potential shortcuts and loops along the way. As can be seen in the SPIKY-flowchart in Fig. \ref{fig:SPIKY-Flowchart}, the general flow is clearly directed from the input of spike train data to the output of results. So the first step the user has to do is to give SPIKY spike train data (i.e. sequences of spike times) to work with. There are four possible ways to do this: one can make use of predefined examples, load data from a file or from the Matlab workspace, detect discrete events from continuous data, or employ the spike train generator (see Section \ref{sss:Input} for more details on the SPIKY input).
%
%
%

\begin{figure}
    \includegraphics[width=85mm]{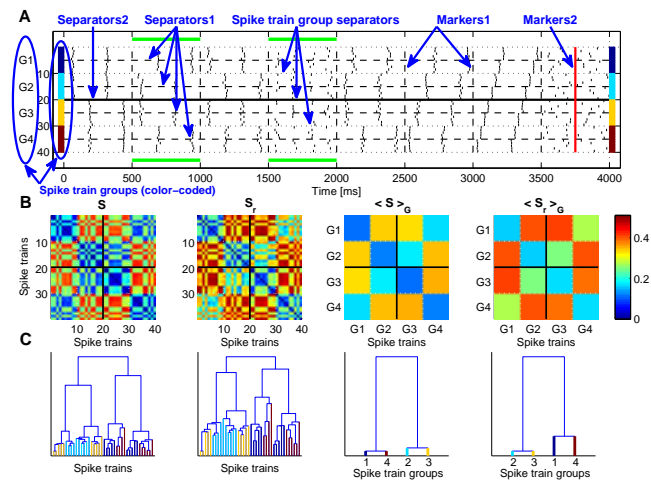}
    \caption{\abb\label{fig:Movie-Screenshot} Annotated screenshot from a movie.   A. Artificially generated spike trains.   B. Dissimilarity matrices obtained by averaging over two separate time intervals for both the regular and the real-time SPIKE-distance as well as their averages over subgroups of spike trains (denoted by $<\cdot>_G$).   C. Corresponding dendrograms.}
\end{figure}

Once the full dataset is available, modification is still possible. The user can restrict the analysis to a specific subset, e.g., select a smaller time window and/or a subset of spike trains. It is also possible to impose some external structure on the raster plot (spike trains vs time). For that, SPIKY allows the definition of two types of time markers (e.g. thick/thin markers for specific events such as seizure onset/offset in epilepsy, trigger onset/offset during stimulation etc.) and two types of spike train separators (e.g. a thick separator for neurons from the left vs. neurons from the right hemisphere and a thin separator for different regions within the two halves). The user can also define spike train groups. Depending on the setup these could be spike trains recorded in different brain regions or upon presentation of different stimuli. Fig. \ref{fig:Movie-Screenshot} shows an example of a raster plot with annotations marking all these different elements.
%
%
%

After updating all of these data parameters, the next step is to select the measures to be calculated. Options include the Peri-Stimulus Time Histogram as well as all measures described in detail in Section \ref{s:Measures}. In the same step the user can select successive frames for a temporal analysis of spike train patterns. These can be individual time instants for cross sections, temporal intervals for selective averages and sequences of time instants for triggered averages (see Section \ref{s:Information-extraction}).

Now the actual calculation of the measures takes place. For reasonably sized datasets this should take at most a few seconds. Very large datasets (typically datasets containing hundreds of spike trains and/or hundred thousands of spikes) are divided in smaller subintervals and the calculation will be performed in a loop which might take longer. It is from this point on that SPIKY becomes truly interactive. Now the user can switch between different representations of the results (such as dissimilarity profiles, dissimilarity matrices and dendrograms). Different matrices and dendrograms can be compared in the same figure or they can be viewed in sequence.

The presentation can be restricted to smaller time windows and/or subsets of spike trains, and temporal and spatial averaging (for example moving average and average over spike train groups) can be performed. Spike trains or spike train groups can also be sorted according to the number of spikes (either within the whole spike train or within a certain interval) or according to the spike latency with respect to a specific time instant. The order can even depend on the result of the clustering analysis, i.e., spike trains belonging to the same cluster will appear next to each other.

Furthermore, it is possible to add further figure elements such as spike number histograms, overall averages, or dissimilarity profiles for individual spike train groups. At this stage the user can also retrospectively change the appearance of all the individual elements of the figure (see Section \ref{sss:Figure-Layout} for more details). Finally, SPIKY allows to extract both data and results to the Matlab workspace for further analysis, and it is also possible to save individual figures as postscript-file or a sequence of figures as an `avi'-movie (see Section \ref{sss:Output} for more details on the SPIKY output).

\subsubsection{\label{sss:Input} Input}

There are four different ways to input spike train data into SPIKY.

The first option is to select one of the predefined examples which are generated using Matlab-code. Initially these are the examples used in \citet{Kreuz13} but one can also define new examples.

The second option is to load spike train data either from the Matlab workspace or from a file. Two different file formats are accepted, `.mat' and `.txt' (ASCII) files. For the mat-files SPIKY currently allows three different kinds of input formats (further formats can be added on demand):

\begin{itemize}
\item cell arrays (ca) with just the spike times. This is the preferred format used by SPIKY since it is most memory efficient. The two other formats will internally be converted into this format;
\item rectangular matrices with each row being a spike train and zero padding (zp) in case of non-identical spike numbers	;
\item matrices representing time bins where each zero/one (01) indicates the absence/presence of a spike.
\end{itemize}

In case of a mat-file or of the workspace, SPIKY looks for a variable called `spikes'; if it cannot find it, the user has the chance to select the variable name (or field name) which contains the spikes via an input mask which provides a hierarchical structure tree of all the variables and fields contained in the mat-file or in the workspace.

In the text format spike times should be written as a matrix with each row being one spike train. The SPIKY-package contains one example file for all four formats (`testdata\_ca.mat', `testdata\_zp.mat', `testdata\_01.mat' and `testdata.txt').

The third option is to use the \textit{event detector} in order to detect discrete events in continuous data. There are many different possibilities of defining an event. A variety of standard events (such as local maxima and minima and threshold crossings) with a number of parameters are already included. 

The fourth option is to create new spike train data via the \textit{spike train generator}. After setting some defining variables (number of spike trains, start and end time, sampling interval) the user can build spike trains from predefined spike train patterns (such as periodic, splay, uniform or Poisson) and/or by manually adding, shifting and deleting individual spikes or groups of spikes.

\subsubsection{\label{sss:Figure-Layout} Figure-Layout}

SPIKY is designed so that it can directly generate figures suitable for publication. The user is given control over the appearance of every individual element (e.g. fonts, lines etc.) in each type of figure. There are two ways to determine essential properties such as color, font size or line width. It is possible to change elements in the active figure while the program is already running. Context menus let the user edit the properties of individual elements or of all elements of a certain type. Conveniently, one can also use the file `SPIKY\_f\_user\_interface' to define the standard values for all the parameters that describe the principal layout of the figure.

If a figure contains more than one subplot (besides the subplot containing the spike rasterplot and/or the dissimilarity profiles, these are typically subplots with dissimilarity matrices and dendrograms), it is possible to change their position and size. One can edit all position variables together or change the x-position, the y-position, the width and the height individually. In case there are several dissimilarity matrices/dendrograms this can be done either for an individual matrix/dendrogram or for all of them at once.

\subsubsection{\label{sss:Output} Output}

From within SPIKY it is possible to extract the spike trains and the results of the analysis (measure profiles, matrices, dendrograms) to the Matlab workspace for further processing. Results will be stored in variables such as `SPIKY\_spikes', `SPIKY\_profile\_X\_1', `SPIKY\_profile\_Y\_1', `SPIKY\_profile\-\_name\_1', `SPIKY\_ma\-trix\_1' and `SPIKY\_matrix\_name\_1'. In addition, the results obtained during an analysis will automatically be stored in the output structure `SPIKY\_results' which will have one field for each measure selected. Depending on the parameter selection within SPIKY, for each measure the structure can contain the following subfields that correspond to the different representations identified in Section \ref{s:Information-extraction}:

\begin{itemize}
\item SPIKY\_results.$<$Measure$>$.name: Name of selected measure
\item SPIKY\_results.$<$Measure$>$.distance: Level of dissimilarity over all spike trains and the whole interval. This is just one value, obtained by averaging over both spike trains and time
\item SPIKY\_results.$<$Measure$>$.matrix: Pairwise distance matrices, obtained by averaging over time
\item SPIKY\_results.$<$Measure$>$.time: Time-values of overall dissimilarity profile
\item SPIKY\_results.$<$Measure$>$.profile: Overall dissimilarity profile obtained by averaging over spike train pairs
\end{itemize}

Note that the dissimilarity profiles are not equidistantly sampled. Rather they are stored as memory-efficiently as possible which means just one value for each interval of the pooled spike train for the ISI- and two values for the SPIKE-distance. Since this format can be more difficult to process, SPIKY includes three functions: `SPIKY\_f\_selective\_averaging' for computing the selective average over time intervals, `SPIKY\_f\_triggered\_averaging' for calculating the triggered average over time instants, and `SPIKY\_f\_average\_pi' for averaging over many dissimilarity profiles. Furthermore, for the ISI-distance the function `SPIKY\_f\_pico' can be used to obtain the average value as well as the x- and y-vectors for plotting.

Besides the standard way to work with Matlab-figures SPIKY also offers the opportunity to save each figure as a postscript-file. Finally, it is possible to save a sequence of figures as an `avi'-movie.

\subsection{\label{ss:GUI-vs-loop} GUI vs. loop}

SPIKY was mainly designed to facilitate the detailed analysis of one dataset. It enables the user to switch between different representations (see Section \ref{s:Information-extraction}) and to zoom in on both spatial and temporal features of interest. However, SPIKY is less convenient for the analysis of many different datasets when e.g. the statistics of a certain quantity such as an average over a specific time interval should be evaluated over all available datasets (e.g. over all trials of a stimulus setup or for recordings of all subjects etc.) in some kind of loop. For these purposes the SPIKY-package contains a program called SPIKY\_loop which is complementary to SPIKY. It is not a graphical user interface but it should be simple enough (and plenty of examples are provided) to allow everyone to run the same kind of analysis for many different datasets and to evaluate and compare their `SPIKY\_results'. SPIKY\_loop provides the full functionality of SPIKY and gives access to time instants, selective and triggered averages as well as averages over spike train groups.

So by combining these two programs it is possible to first use SPIKY for a rather exploratory but detailed analysis of a limited number of individual datasets and then use SPIKY\_loop and its output structure `SPIKY\_loop\_results' to verify whether any effect discovered on the example dataset is consistently present within all of the datasets.

Both SPIKY and SPIKY\_loop require the storage of matrices of the order `number of interspike intervals in the pooled spike train' $\times$ `number of spike train pairs'. For very large datasets with many spike trains and/or spikes this can lead to memory problems. We addressed this issue by making the calculation sequential, i.e., by cutting the recording interval into smaller segments, and performing the averaging over all pairs of spike trains for each segment separately. In the end the dissimilarity profiles for the different segments (already averaged over pairs of spike trains) are concatenated, and its temporal average yields the distance value for the whole recording interval. During this sequential calculation SPIKY uses a waitbar (which displays the \% completed) to continuously inform the user about the progress. This way, by trading memory against speed - running more loops with smaller matrices takes longer - SPIKY is able to deal with datasets containing more than one hundred spike trains and overall more than one million spikes.

\subsection{\label{ss:Spike-train-surrogates} Spike train surrogates and significance}

A very important issue that has not yet been addressed is statistical significance. Given a certain value of the SPIKE-distance how can one judge whether it reflects a significant decrease or increase in spike train synchrony and does not just lie within the range of values obtained for random fluctuations? One answer to this question is the use of spike train surrogates \citep{Kass05, Gruen09, Louis10}. The idea is to compare the results for the original dataset with the results obtained for spike train surrogates generated from that dataset. If the value obtained for the original lies outside the range of values for the surrogates this value can be assumed to be significant to a level defined by the number of surrogates used (e.g. $\alpha = 0.05$ for $19$ surrogates or $\alpha = 0.001$ for $999$ surrogates).

The SPIKY-package contains a program `Spiky\_loop\_\-surro' which was designed to evaluate significance. So far it includes four different types of spike train surrogates. They differ in the properties that are preserved and maintain either the individual spike numbers (obtained by shuffling the spikes), the individual interspike interval distribution (obtained by shuffling the interspike intervals), the pooled spike train (obtained by shuffling spikes among the spike trains) or the Peri-Stimulus Time Histogram (PSTH). In the last case the spike train surrogates are obtained by means of inverse transform sampling, i.e., by resampling from the PSTH using its cumulative distribution function (CDF) \citep{Ross97}. Other ways to calculate rate functions (e.g. based on kernels with different bandwidths) will be added in future releases.

%
%
\section{\label{s:Discussion} Discussion}

\subsection{\label{ss:Summary} Summary}

In this paper we presented SPIKY, a graphical user interface which facilitates the application of methods of spike train analysis to both simulated and real datasets. Apart from the standard Peri-Stimulus Time Histogram, SPIKY contains three parameter-free and time-resolved measures (see Section \ref{s:Information-extraction}). These measures are complementary to each other since each one addresses a different specific aspect of spike train similarity. While the ISI-distance quantifies local dissimilarities based on covariances of the neurons’ firing rate profiles, both the SPIKE-distance and SPIKE-synchronization capture the relative timing of local spikes\footnote{One should be aware that all of these measures are based on some explicit notion of localness and simultaneousness and thus they all can be fooled by phase lags, no matter whether these are caused by internal delay loops in one spike train or by a common driver that forces the two time series with different delays. Therefore, any such phase lags should be removed by suitably shifting the time series in a preprocessing step.}. However, whereas the SPIKE-distance weights and normalizes the differences between nearest neighbor spikes, SPIKE-synchronization acts as a binary coincidence detector, i.e. there is a cutoff at the (adaptive) time lag relative to which two neighboring spikes are either considered coincident or not and all detailed information both within or outside this coincidence window is discarded.

All of these measures yield instantaneous values for each pair of spike trains, and thus there are many different possible representations of the results (see Section \ref{s:Information-extraction}). Often the most informative representation might depend on the amount and type of spike train data and SPIKY can be used to reveal it via some explorative and interactive analysis. SPIKY also allows to alter a given dataset before or after the actual analysis, e.g., to interactively select subintervals or spike train subsets, to define time markers and spike train separators, and to divide the dataset into different spike train groups. 

In addition to the main GUI designed for the detailed analysis of one dataset, the SPIKY-package also includes two complementary programs. While SPIKY\_loop aims at the grand average analysis of large numbers of datasets, SPIKY\_loop\_surro allows the estimation of significance levels. In all of these programs we use MEX-files, i.e. C-based Matlab executables for the more time-consuming parts and we exploit the piecewise linearity of the dissimilarity profiles, thereby guaranteeing high computation speed and memory efficiency.

\subsection{\label{ss:Outlook} Outlook}

One of the measures included in SPIKY is the realtime SPIKE-distance. The present algorithm calculates its instantaneous dissimilarity values by making use of past information only but it does so in a speed-optimized and parallelized manner which would not be compatible with an actual realtime-implementation. However, such a realtime-implementation should actually be simpler and computationally (speed and memory) less problematic even for large numbers of neurons since the only information that would have to be stored at each time instant are the time stamps of the latest spikes of each spike train and their nearest neighbors in the other spike trains.

SPIKY was primarily designed to analyze neuronal spike trains. But in principle it is also applicable to any other kind of discrete data which comes in the form of sequences of time stamps (such as times of bouncing basketballs or time-coded social interactions in a psychological test). Furthermore, in \citet{Kreuz13} the SPIKE-distance was already applied not only to discrete data but also to an example of continuous data (EEG). To clear the way for such applications SPIKY includes an event detector which allows to bridge the gap between continuous data and our methods for discrete data.

We chose to write SPIKY in Matlab due to its popularity in the neuroscience community, ease of use, and the engaging visualization capabilities of its GUI design. Because of its high level parallelization, Matlab provides a powerful tool for processing vectorized data, but it also includes a well-developed MEX-interface for integrating C functions for performance enhancements. C functions do not only lead to an increase in performance, they can also easily be incorporated into other programming platforms. Indeed, we have already ported all four measures (the PSTH, the ISI-distance, the SPIKE-distance and SPIKE synchronization) as well as some of the additional SPIKY-functionality to Python as part of the PySpike\footnote{https://github.com/mariomulansky/PySpike} module, an open-source project hosted on Github. As in SPIKY, in PySpike the computation intensive functions are implemented as C backends.

Another potential direction would be the extension of the methods used here from the quantification of synchrony within one population of two or more neurons to the quantification of synchrony between two (or more) neuronal populations. Similar extensions have been done for two other spike train distances \citep{Aronov03, Houghton08} but both of these methods depend on not one but two parameters. So one particular challenge for a potential extension of our methods would be to make them parameter-free while maintaining their high temporal resolution.

There are still many open questions regarding neuronal synchrony. Among these questions are its role in dynamical diseases like epilepsy and its relevance for neural coding. While a thorough discussion of these issues is beyond the scope of this paper, we argue that SPIKY will be a very useful tool for investigation. If epileptic seizures and/or time-dependent stimulations lead to changes in spike train synchrony or spike train clustering, SPIKY should be able to detect them.

\vspace{1cm}

\begin{thanks}
\section{\label{s:Acknowledgement} \textbf{Acknowledgements}}

TK, MM, and NB acknowledge funding support from the European Commission through the Marie Curie Initial Training Network `Neural Engineering Transformative Technologies (NETT)', project 289146, and TK also through the European Joint Doctorate 'Complex oscillatory systems: Modeling and Analysis (COSMOS)', project 642563. TK acknowledges the Italian Ministry of Foreign Affairs regarding the activity of the Joint Italian-Israeli Laboratory on Neuroscience and NB acknowledges the Serbian Ministry of Youth and Sports. TK thanks Marcus Kaiser and his group for hosting him at the University of Newcastle, UK. NB thanks John van Opstal and his group for hosting him at the Radboud University, Nijmegen, Netherlands as well as Joshua D Berke and his group for hosting him at the University of Michigan, Ann Arbor, MI, USA.
     
We thank Ralph Andrzejak, David Angulo Garcia, Francesco Battaglia, Roman Bauer, Joshua D. Berke, Paolo Bonifazi, Emily Caporello, Daniel Chicharro, Michael Farries, Tim Gentner, Bon-Mi Gu, Arif Hamid, Conor Houghton, Marcus Kaiser, Jutta Kretzberg, Tatjana Loncar Turukalo, Stefano Luccioli, Jason MacLean, Gorana Mijatovic, Ali Mohebi, Florian Mormann, Leon Paz, Jeffrey Pettibone, Friederice Pirschel, Andreea Sburlea, Peter Taylor, Richard Tomsett, Alessandro Torcini, Jonathan Victor, and Yujiang Wang for useful discussions.

We also thank Thomas Alderson, Mayte Bonilla Quintana, Hamid Charkhkar, Didier Desaintjan, Mario DiPoppa, Andres Espinal Jim\'{e}nez, Mahboubeh Etemadi, Gabriel Chew Guojun, Taekyung Kang, Marion Najac, Oren, Robert Rein, Rodrigo Salazar, Andreea Sburlea, Michael Schaub, Eitan Schechtman, Jannetta Steyn, Matthew Williams, Yunguo Yu, Maja Zorovic for user feedback and Black Square, Conor Houghton, Eugenio Piasini, Matthew Phillips, and Sid Visser for advice.

Finally, we thank Daniel Chicharro, Conor Houghton, and Andreea Sburlea for carefully reading the manuscript.

\end{thanks}

\bibliographystyle{elsart-harv}


\begin{thebibliography}{25}
 
 \bibitem[{Aronov(2003)}]{Aronov03}
 Aronov D, Reich DS, Mechler F, Victor JD. Neural coding of spatial phase in V1 of the macaque monkey. J Neurophysiol, 2003;89:3304-27.
 
 \bibitem[{Bower et al.(2012)}]{Bower12}
 Bower MR, Stead M, Meyer FB, Marsh WR, Worrell GA. Spatiotemporal neuronal correlates of seizure generation in focal epilepsy. Epilepsia, 2012;53:807-16.
 `
 \bibitem[{{Di Poppa} and Gutkin(2013)}]{DiPoppa13}
 {DiPoppa} M, Gutkin BS. Correlations in background activity control persistent state stability and allow execution of working memory tasks. Front Comp Neurosci, 2013;7:139.
 
 \bibitem[{Gruen(2009)}]{Gruen09}
 Gr{\"u}n S. Data-Driven Significance Estimation for Precise Spike Correlation. J Neurophysiol, 2009;101:1126-40.
 
 \bibitem[{Hochberg et al.(2006)}]{Hochberg06}
 Hochberg LR, Serruya MD, Friehs GM, Mukand JA, Saleh M, Caplan AH, Branner A, Chen D, Penn RD, Donoghue JP. Neuronal ensemble control of prosthetic devices by a human with tetraplegia. Nature, 2006;442:164-71.
 
 \bibitem[{Houghton and Sen(2008)}]{Houghton08}
 Houghton C, Sen K. A new multineuron spike train metric. Neural Comput, 2008;20:1495-511.
 
 \bibitem[{Kass et al.(2005)}]{Kass05}
 Kass RS, Ventura V, Brown EN. Statistical issues in the Analysis of Neuronal Data. J Neurophysiol, 2005;94:8-25.
 
 \bibitem[{Kreuz et al.(2007)}]{Kreuz07c}
 Kreuz T, Haas JS, Morelli A, Abarbanel HDI, Politi A. Measuring spike train synchrony. J Neurosci Methods, 2007;165:151-61.
 
 \bibitem[{Kreuz et al.(2009)}]{Kreuz09}
 Kreuz T, Chicharro D, Andrzejak RG, Haas JS, Abarbanel HDI. Measuring multiple spike train synchrony. J Neurosci Methods, 2009;183:287-99.
 
 \bibitem[{Kreuz et al.(2011)}]{Kreuz11}
 Kreuz T, Chicharro D, Greschner M, Andrzejak RG. Time-resolved and time-scale adaptive measures of spike train synchrony. J Neurosci Methods, 2011;195:92-106.
 
 \bibitem[{Kreuz(2011)}]{Kreuz11b}
 Kreuz T. Measures of spike train synchrony. Scholarpedia, 2011;6:11934.
 
 \bibitem[{Kreuz(2012)}]{Kreuz12}
 Kreuz T. SPIKE-distance. Scholarpedia, 2012;7:30652.
 
 \bibitem[{Kreuz et al.(2013)}]{Kreuz13}
 Kreuz T, Chicharro D, Houghton C, Andrzejak RG, Mormann F. Monitoring spike train synchrony. J Neurophysiol, 2013;109:1457.
 
 \bibitem[{Kumar et al.(2010)}]{Kumar10}
 Kumar A, Rotter S, Aertsen A. Spiking activity propagation in neuronal networks: reconciling different perspectives on neural coding. Nature Rev Neurosci, 2010;11:615-27.
 
 \bibitem[{Louis et al.(2010)}]{Louis10}
 Louis S, Gerstein GL, Gr{\"u}n S, Diesmann M. Surrogate spike train generation through dithering in operational time. Front Comp Neurosci, 2010;4:127.
 
 \bibitem[{Mainen and Sejnowski(1995)}]{Mainen95}
 Mainen Z, Sejnowski TJ. Reliability of spike timing in neocortical neurons. Science 1995;268:1503-6.
 
 \bibitem[{Miller and Wilson(2008)}]{Miller08}
 Miller EK, Wilson MA. All My Circuits: Using Multiple Electrodes to Understand Functioning Neural Networks. Neuron, 2008;60:483-8.
 
 \bibitem[{Mormann et~al.(2007)}]{Mormann07}
 Mormann F, Andrzejak RG, Elger CE, Lehnertz K. Seizure prediction: the long and winding road. Brain, 2007;130:314-33.
 
 \bibitem[{Nirenberg and Victor(2007)}]{Nirenberg07}
 Nirenberg S, Victor JD. Analyzing the activity of large populations of neurons: how tractable is the problem? Curr Opin Neurobiol 2007;17:397-400.
 
 \bibitem[{Papoutsi et~al.(2013)}]{Papoutsi13}
 Papoutsi A, Sidiropoulou K, Cutsuridis V, Poirazi P. Induction and modulation of persistent activity in a layer v pfc microcircuit model. Front Neural Circuits 2013;7:161.
 
 \bibitem[{Quian Quiroga} et~al.(2002)]{QuianQuiroga02b}
 {Quian Quiroga} R, Kreuz T, Grassberger P. Event synchronization: \textsc{A} simple and fast method to measure synchronicity and time delay patterns. Phys Rev E 2002;66:041904.
 
 \bibitem[{Quian Quiroga} and Panzeri(2013)]{QuianQuiroga13}
 {Quian Quiroga} R, Panzeri S (Eds.). Principles of neural coding. CRC Taylor and Francis, Boca Raton, FL, USA 2013.
 
 \bibitem[{Ross(1997)}]{Ross97}
 Ross SM. Introduction to Probability Models. Sixth edition. Academic Press, New York, USA 1997.
 
 \bibitem[{Rusu and Florian(2014)}]{Rusu14}
 Rusu CV, RV Florian. A new class of metrics for spike trains. Neural Computation, 2014;26:306-48.
 
 \bibitem[{Sacr\'{e} and Sepulchre(2014)}]{Sacre14}
 Sacr\'{e} P, Sepulchre R. Sensitivity analysis of oscillator models in the space of phase-response curves: Oscillators as open systems. Control Systems, IEEE 2014;34:50–74.
 
 \bibitem[{Sanchez(2008)}]{Sanchez08}
 Sanchez JC, Principe JC, Nishida T, Bashirullah R, Harris JG, Fortes JAB. Technology and Signal Processing for Brain-Machine Interfaces. IEEE Signal Processing, 2008;25:29-40.
 
 \bibitem[{Shlens et al.(2008)}]{Shlens08}
 Shlens J, Rieke F, Chichilnisky EJ. Synchronized firing in the retina. Curr Opin Neurobiol, 2008;18:396-402.
 
 \bibitem[{Thibeault et al.(2014)}]{Thibeault14}
 Thibeault CM, {O'Brien} MJ, Srinivasa N. Analyzing large-scale spiking neural data with HRLAnalysis. Front. Neuroinform., 2014;8:17.
 
 \bibitem[{Tiesinga et al.(2008)}]{Tiesinga08}
 Tiesinga PHE, Fellous JM, Sejnowski TJ. Regulation of spike timing in visual cortical circuits. Nature Reviews Neuroscience 2008;9:97-107.
 
 \bibitem[{Truccolo et al.(2011)}]{Truccolo11}
 Truccolo W, Donoghue JP, Hochberg LR, Eskandar EN, Madsen JR, Anderson WS, Brown EN, Halgren E, Cash SS. Single-neuron dynamics in human focal epilepsy. Nature Neurosci, 2011;14:635-41.
 
 \bibitem[{Victor(2005)}]{Victor05}
 Victor JD. Spike train metrics. Current Opinion in Neurobiology 2005;15:585-592.
 
 \end{thebibliography}


\end{document}